\documentclass[runningheads,a4paper]{llncs}

\usepackage{amssymb}
\usepackage{graphicx}
\usepackage{tabularx}
\usepackage{tabulary}
\usepackage{booktabs}
\usepackage{epstopdf}
\usepackage{float}
\usepackage[hyphens]{url}
\usepackage{hyperref}
\usepackage[all]{hypcap}
\usepackage{tabto}
\usepackage{amsmath}
\usepackage{enumitem}
\usepackage[autostyle]{csquotes}
\usepackage{listings}
\usepackage{subfig}
\usepackage{color}

\usepackage[numbers,square]{natbib}

\newcommand{\keywords}[1]{\par\addvspace\baselineskip
\noindent\keywordname\enspace\ignorespaces#1}

\begin{document}

\mainmatter

\title{Linking Mathematical Software\\in Web Archives\thanks{This work is partly funded by the German Research Council under FID Math and the European Research Council under ALEXANDRIA (ERC 339233)}}

\titlerunning{Linking Mathematical Software in Web Archives}

\author{Helge Holzmann\inst{1} \and  Mila Runnwerth\inst{2} \and Wolfram Sperber\inst{3}}

\authorrunning{Holzmann et al.}

\institute{L3S Research Center\\
Appelstr. 9a, 30167 Hannover, Germany\\\email{holzmann@L3S.de}\\
\and
German National Library of Science and Technology (TIB)\\
Welfengarten 1b, 30167 Hannover, Germany\\\email{Mila.Runnwerth@tib.eu}\\
\and
zbMATH, FIZ Karlsruhe - Leibniz Institute for Information Infrastructure\\
Franklinstr. 11, 10587 Berlin, Germany\\\email{wolfram@zentralblatt-math.org}}

\maketitle

\begin{abstract}
The Web is our primary source of all kinds of information today. This includes information about software as well as associated materials, like source code, documentation, related publications and change logs. Such data is of particular importance in research in order to conduct, comprehend and reconstruct scientific experiments that involve software. \textsf{swMATH}, a mathematical software directory, attempts to identify software mentions in scientific articles and provides additional information as well as links to the Web. However, just like software itself, the Web is dynamic and most likely the information on the Web has changed since it was referenced in a scientific publication. Therefore, it is crucial to preserve the resources of a software on the Web to capture its states over time.

We found that around 40\% of the websites in \textsf{swMATH} are already included in an existing Web archive. Out of these, 60\% of contain some kind of documentation and around 45\% even provide downloads of software artifacts. Hence, already today links can be established based on the publication dates of corresponding articles. The contained data enable enriching existing information with a temporal dimension. In the future, specialized infrastructure will improve the coverage of software resources and allow explicit references in scientific publications.

\end{abstract}

\section{Introduction}
\label{sec:introduction}

Providing specialized information services for software (SW) is challenging for various reasons: SW is highly dynamic, references in literature are often not declared as SW, and structured metadata is sparse. Repositories as well as directories typically represent SW in an abstract manner with provided information corresponding not to a specific state or only to the current version. In contrast to these representations, mentions of SW in scientific articles often refers to the SW's state at the time of use or publication. However, SW is dynamic and often a version has been updated since it was references. With changes of the SW, its website is likely to be updated as well. Thus, corresponding information as well as associated materials of the referenced version might not available anymore. This makes it difficult to trace the development process and to obtain information about a previous version of the SW, although this can be necessary to reproducing published scientific results. While open source SW facilitates access to different states of the source code, this does not always include associated materials, such as a documentation.

To fix this temporal gap between publications and representations of SW on the Web in the long run, a sophisticated Web archiving infrastructure is required to preserve available information about SW at the time when it is referenced. In the short term, however, establishing links to existing Web archives at the publication time of an article will be a first step in this direction. SW directories can serve as a bridge in this endeavor. \textsf{swMATH}\footnote{\url{http://www.swmath.org}}, a directory for mathematical SW (MSW), follows a publication-based approach to link SW and corresponding scientific publications (s. Sec.~\ref{sec:swmath}). From all MSW listed on \textsf{swMATH}, we found the websites of around 40\% to be already existent in the \textit{Internet Archive}'s Web archive\footnote{\url{http://web.archive.org}}, with many of them providing associated materials (s. Sec.~\ref{sec:swmath}).

\section{Publication-based approached of \textsf{swMATH}}
\label{sec:swmath}

\begin{figure}[t]
	\centering
	\includegraphics[width=\textwidth]{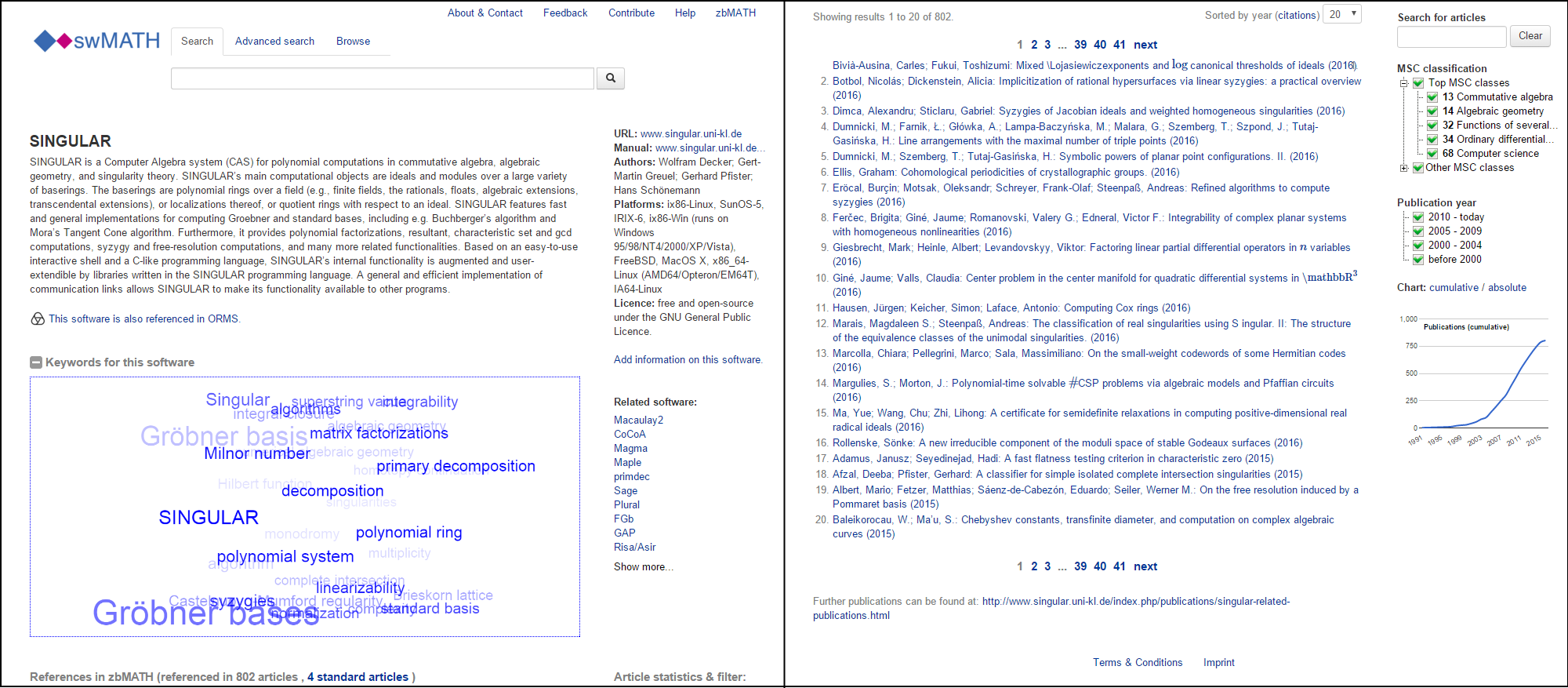}
	\caption{Record of the mathematical Software \textsf{Singular} on \textsf{swMath}}
	\label{fig:swMath}
	\vspace{-15pt}
\end{figure}

\textsf{swMATH} is one of the most comprehensive information services for MSW \citep{greuel2014swmath}. Based on simple heuristics, \textsf{swMATH} identifies MSW in scientific articles from its underlying bibliographic database \textsf{zbMATH}\footnote{\url{http://www.zbmath.org}}, consisting of nearly 120,000 publications referring to MSW. Currently, \textsf{swMATH} contains more than 12,500 SW records linked to corresponding articles, as shown in Figure~\ref{fig:swMath}.

One of the major challenges of \textsf{swMATH} is the identification of MSW in scientific publications. In many articles, only names are mentioned, while versions or an explicit labels as MSW are missing. \textsf{swMATH} addresses this by scanning titles, abstracts, as well as references for typical terms, such as \textit{solver}, \textit{program}, or simply \textit{software}, in combination with a name. After that, a manual intervention step verifies the recognized SW to ensure a high quality of the service. As part of this, additional metadata, such as a website, the authors, technical requirements, dependencies, licenses, documentations and more, is looked-up through a regular Web search and added if available.

Similar to other SW directories and repositories (s. Sec.~\ref{sec:introduction}), the focus of \textsf{swMATH} is to provide time-agnostic information about SW rather than specific versions. Therefore, included websites are periodically checked and outdated links as well as related information are removed or replaced. While this ensures an up-to-date representation of the SW, it introduces inconsistencies with included publications, which are annotated with the year of publication and constitute temporal witnesses of different SW states. Therefore, it has been considered to integrate temporal information in order to represent the different versions of a SW over time and match publications. Web archives can serve as source for this information in the future.

\section{Linking Web Archives}
\label{sec:approach}

Web archives have recently been of growing interest in research, however, they have either been used as scholarly sources or have been a subject of research themselves, with questions focusing on coverage and evolution \citep{ainsworth2011much, holzmann2016jcdl}. To the best of our knowledge, Web archives have never been used to recover information and associated materials of former states of SW. We tackle this by linking articles on SW with available resources in Web archives. As proxy serve the website URLs as well as the publications dates of corresponding articles as listed on \textsf{swMATH}. An initial analysis of this data unveiled what to expect from integrating Web archives as source for temporal information about MSW.

\begin{figure}[t!]
	\centering

	\captionsetup[subfigure]{oneside,margin={0cm,0.7cm}}

	\subfloat[Information on mathematical software pages.]{
		\hspace{-25pt}
		\includegraphics[width=0.55\textwidth]{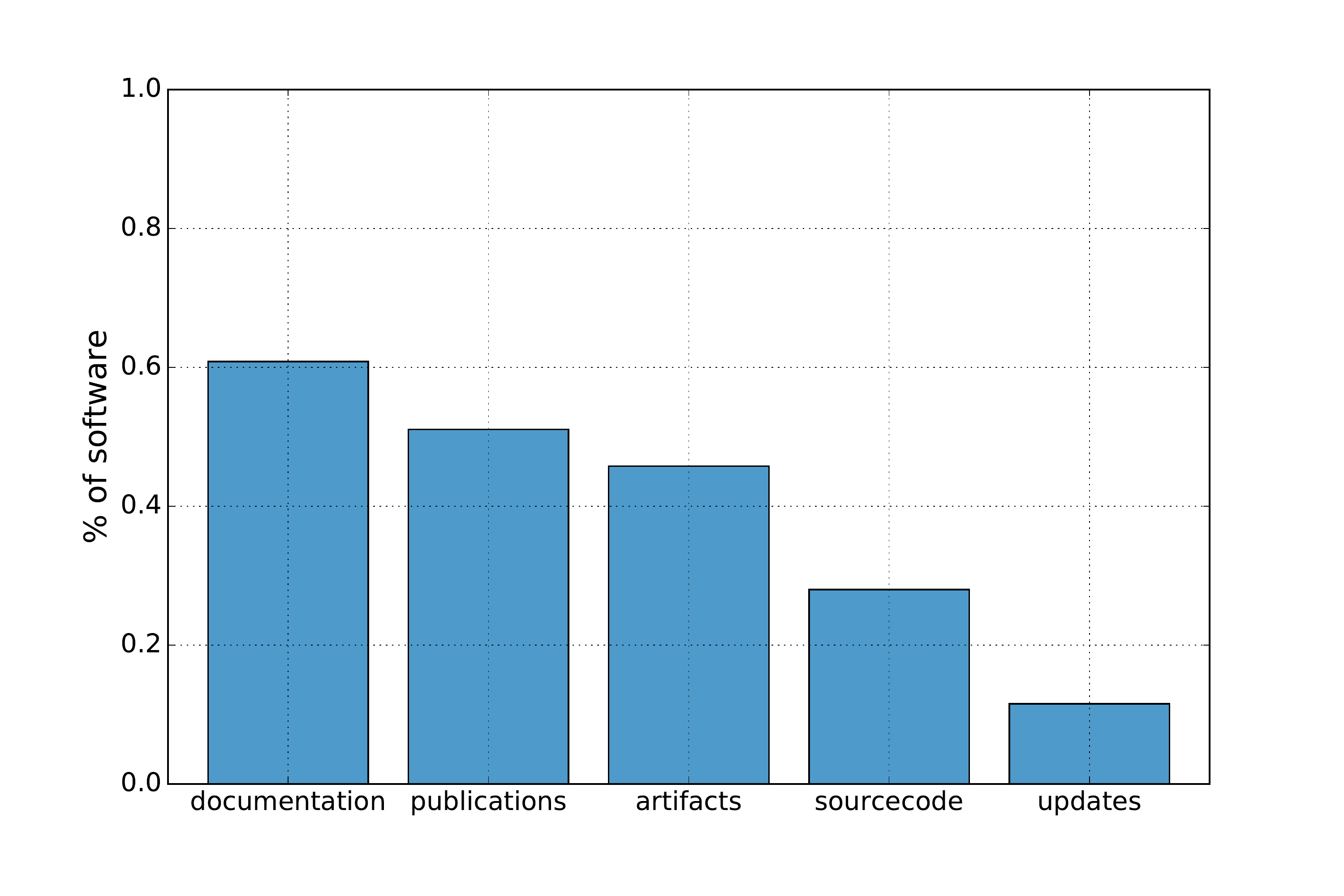}
		\label{fig:information}
	}
	\subfloat[Pages changed since top publication.]{
		\hspace{-12pt}
		\includegraphics[width=0.55\textwidth]{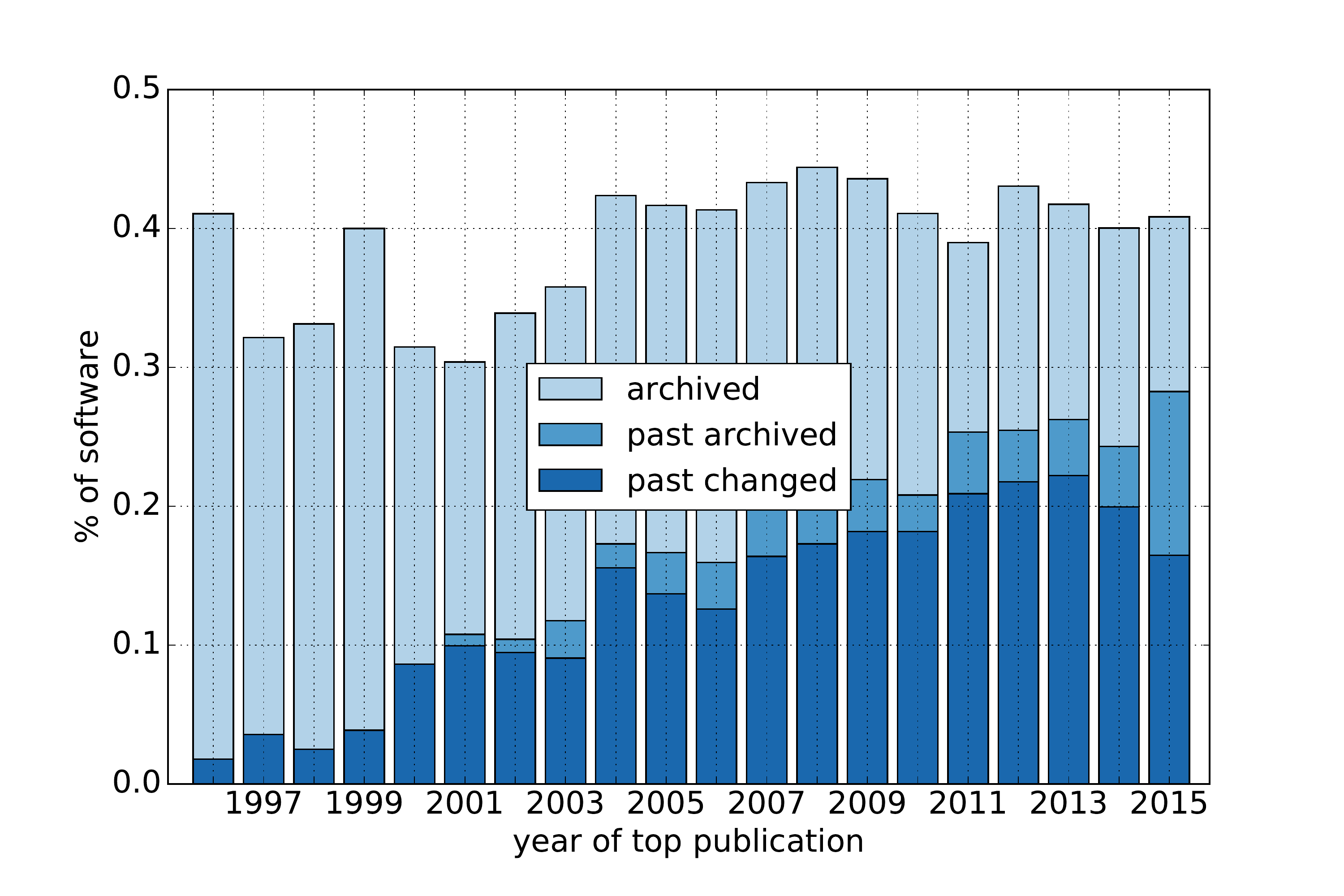}
		\label{fig:archived}
	}

	\caption{Mathematical Software in Web archives.}
	\vspace{-15pt}
\end{figure}

As shown in Figure~\ref{fig:information}, we found that around 60\% of the analyzed websites contain some kind of documentation, almost 50\% link to publications, more than 40\% even provide downloads of SW artifacts with 30\% being open source, and 10\% could be identified to publish updates or news, such as a changelog. Although not all of this is currently being preserved by Web archives, already today around 40\% of the URLs under investigation are included in the considered Web archive with at least one capture, as shown by the \textit{archived} bars in Figure~\ref{fig:archived} dissected by the year of the highest cited publication of a SW. However, looking at the fraction with captures in the corresponding year, the \textit{past archived} bars unveil a clear growth over time with relatively low numbers in the early times of Web archiving. Hence, we are able to successfully link around 25\% of the analyzed MSW with their top publication in 2013, but only very few before 2000. The reason for this is two-fold, while the coverage of Web archives has drastically improved over time, as pointed out in Section~\ref{sec:swmath}, outdated links in \textsf{swMATH} have been replaced with new URLs, which might not have existed at the time of the top publication yet.

\section{Conclusion and Outlook}

Linking MSW in Web archives will help to recover information as well as associated materials of previous versions referenced in scientific publications. As shown by the third category in Figure~\ref{fig:archived}, \textit{past changed}, almost all of the websites that were archived in the year of the top publication mentioning a SW have changed, which indicates the need of our approach. Moreover, the tools used for the analysis of Web archives are a first step towards a machine-based content analysis of the websites of a MSW. This opens up new possibilities to enrich the information in \textsf{swMATH}.

In order to overcome the challenge of identifying former URLs of SW resources in a Web archive (s. Sec.~\ref{sec:approach}), temporal tags may be incorporated in the future, as demonstrated by \citet{holzmann2016www}.

\bibliographystyle{IEEEtranN}
\bibliography{references}

\end{document}